\documentclass{PoS}
\usepackage{amsfonts}
\usepackage{amsmath}
\usepackage{subfigure} 
\usepackage{verbatim}

\graphicspath{{./figs/}}

\newcommand{\MSbar}{\overline{\mathsf{MS}}}

\newcommand{\GeV}{\mathop{\rm GeV}\nolimits}

\title{
  Calculation of BSM Kaon B-parameters using Staggered Quarks 
}

\ShortTitle{
  Calculation of BSM Kaon B-parameters
}
\author{ Yong-Chull Jang, 
  Hwancheol Jeong, Jangho Kim, 
  Seonghee Kim,  Weonjong Lee, \speaker{Jaehoon Leem}, 
  Jeonghwan Pak, Sungwoo Park \\
  Lattice Gauge Theory Research Center, CTP, and FPRD, \\
  Department of Physics and Astronomy,
  Seoul National University, Seoul, 151-747, South Korea \\
  E-mail: \email{wlee@snu.ac.kr}}

\author{Chulwoo Jung, Hyung-Jin Kim \\
  Physics Department, Brookhaven National Laboratory,
  Upton, NY11973, USA \\
  E-mail: \email{chulwoo@bnl.gov}}

\author{Stephen R. Sharpe\\
  Physics Department, University of Washington, 
  Seattle, WA 98195-1560, USA \\
  E-mail: \email{sharpe@phys.washington.edu}}

\author{Boram Yoon \\
  Los Alamos National Laboratory, \\
  Theoretical Division T-2, MS B283, \\
  Los Alamos, NM 87545, USA \\
  E-mail: \email{googlus@gmail.com}}

\author{SWME Collaboration}

\abstract{We present updated results for kaon B-parameters 
for operators arising in models of new physics.
We use HYP-smeared staggered quarks 
on the $N_f = 2+1$ MILC asqtad lattices.
During the last year we have added new ensembles,
which has necessitated chiral-continuum fitting with more elaborate 
fitting functions. We have also
corrected an error in a two-loop anomalous dimension used
to evolve results between different scales.
Our results for the beyond-the-Standard-Model B-parameters
have total errors of $5-10$\%.
We find that the discrepancy observed last year between
our results and those of the RBC/UKQCD and ETM collaborations 
for some of the B-parameters  has been reduced
from $4\!-\!5\,\sigma$ to $2\!-\!3\,\sigma$.
}

\FullConference{ 32nd International Symposium on Lattice Field Theory
  - LATTICE 2014\\ June 23 - June 28, 2014\\ Columbia University }

\begin{document}
\section{Introduction} 
%
In the Standard Model (SM), CP violation in $K-\bar K$ mixing
is proportional to the hadronic matrix element of a left-handed
four-quark operator. This matrix element is conventionally parametrized by
the B-parameter $B_K$.
Beyond the Standard Model (BSM) physics introduces
four additional $\Delta S = 2$ four-quark operators with
different chirality structures.
To constrain BSM models it is thus necessary to have accurate
calculations of the matrix elements of these new operators.
These matrix elements are parametrized by the so-called
BSM kaon B-parameters. 
%

In this talk we present an update on our previous
results for the BSM B-parameters~\cite{Bae:2013mja,Bae:2013tca},
and compare to those from other collaborations using different
fermion discretizations~\cite{Boyle:2012qb,Bertone:2012cu}.
%

%
\section{Methodology}
%

We use the chiral basis of Ref.~\cite{Buras:2000if}
(see Refs.~\cite{Bae:2013mja,Bae:2013tca} for further details):
\begin{align}
{Q}_{1} &= [\bar{s}^a \gamma_\mu (1-\gamma_5) d^a ] 
           [ \bar{s}^b \gamma_\mu (1-\gamma_5) d^b]\,, \nonumber\\
{Q}_{2} &= [\bar{s}^a (1-\gamma_5) d^a]
           [\bar{s}^b (1-\gamma_5) d^b]\,,  \nonumber \\
{Q}_{3} &= [\bar{s}^a \sigma_{\mu\nu}(1-\gamma_5) d^a]
           [\bar{s}^b \sigma_{\mu\nu} (1-\gamma_5) d^b]\,, \\
{Q}_{4} &= [\bar{s}^a (1-\gamma_5) d^a] 
           [\bar{s}^b (1+\gamma_5) d^b]\,, \nonumber \\
{Q}_{5} &= [\bar{s}^a \gamma_\mu (1-\gamma_5) d^a]
           [\bar{s}^b \gamma_\mu (1+\gamma_5) d^b]\,, \nonumber
\end{align}
where $a,b$ are color indices. 
The operator $Q_1$ appears in the SM and its matrix element is parametrized
by $B_K$.
The BSM B-parameters are defined as
\begin{equation}
B_i = \frac{\langle \bar{K}_0 \vert Q_i \vert K_0 \rangle}
{N_i \langle \bar{K}_0 \vert \bar{s}\gamma_5 d\vert 0 \rangle
\langle 0 \vert \bar{s} \gamma_5 d \vert K_0 \rangle}\,,
\label{eq:BSMBdef}
\end{equation}
where $i=2-5$ and $(N_2, N_3, N_4, N_5) = (5/3, 4, -2, 4/3 )$.
Other lattice calculations have used instead
the ``SUSY'' basis of Ref.~\cite{Gabbiani:1996hi}.
The relation of the B-parameters in the two bases is
\begin{equation}
B^{\text{SUSY}}_2 = B_2,\quad
B^{\text{SUSY}}_3 = -\frac{3}{2}B_3  + \frac{5}{2}B_2,\quad
B^{\text{SUSY}}_4 = B_4,\quad
B^{\text{SUSY}}_5 = B_5.
\end{equation}
%

We use the MILC asqtad lattices listed in Table~\ref{tab:milc-lat},
with HYP-smeared staggered valence quarks.
Since Lattice 2013 we have added four new ensembles:
F6, F7, F9, and S5. This allows for more careful 
chiral and continuum extrapolations.
Our data analysis follows almost the same methodology as
previously (see Ref.~\cite{Bae:2013tca}), with some changes
described below.
In particular we continue to use one-loop perturbative matching 
to obtain operators defined in the
continuum $\MSbar$ scheme using naive dimensional regularization.
\begin{table}
\caption{MILC asqtad ensembles. Here ``ens'' is the 
number of gauge configurations, ``meas'' is the number 
of measurements per configuration, and ID is name of 
ensemble. ``New'' ensembles have been added since 
Lattice 2013~\cite{Bae:2013mja}.}
\label{tab:milc-lat}
\begin{center}
\begin{tabular}{|l | l | l | l| l|c |}
\hline\hline
$a$ (fm) & $am_l/am_s$ & geometry & ens$\times$meas & ID &Status\\
\hline
0.09 & 0.0062/0.0310 & $28^3 \times 96$ & $995 \times 9$ &F1& \\
0.09 & 0.0031/0.0310 & $40^3 \times 96$ & $959 \times 9$ &F2& \\
0.09 & 0.0093/0.0310 & $28^3 \times 96$ & $949 \times 9$ &F3& \\
0.09 & 0.0124/0.0310 & $28^3 \times 96$ & $1995 \times 9$&F4& \\
0.09 & 0.00465/0.0310 & $32^3 \times 96$ & $651 \times 9$&F5& \\
0.09 & 0.0062/0.0186 & $28^3 \times 96$ & $950 \times 9$ &F6&New \\
0.09 & 0.0031/0.0186 & $40^3 \times 96$ & $701 \times 9$ &F7&New \\
0.09 & 0.00155/0.0310 & $64^3 \times 96$ & $790 \times 9$&F9&New \\
\hline
0.06 & 0.0036/0.018 & $48^3 \times 144$ & $749 \times 9$ &S1& \\
0.06 & 0.0025/0.018 & $56^3 \times 144$ & $799 \times 9$ &S2& \\
0.06 & 0.0072/0.018 & $48^3 \times 144$ & $593 \times 9$ &S3& \\
0.06 & 0.0054/0.018 & $48^3 \times 144$ & $582 \times 9$ &S4& \\
0.06 & 0.0018/0.018 & $64^3 \times 144$ & $572 \times 9$ &S5&New \\
\hline
0.045 & 0.0030/0.015 & $64^3 \times 192$ & $747 \times 1$&U1& \\
\hline\hline
\end{tabular}
\end{center}
\end{table}

Valence $d$ and $s$ quarks are denoted $x$ and $y$, respectively.
Thus we must extrapolate $m_x$ to $m_d^{\rm phys}$ and
$m_y$ to $m_s^{\rm phys}$. We do the former extrapolation using
next-to-leading order (NLO)
SU(2) staggered chiral perturbation theory (SChPT),
which requires $m_x \ll m_y \approx m_s$.
In practice we use valence masses of $n \times m_s/10$, with
$m_s$ a nominal strange quark mass which depends on the
ensemble and turns out to be somewhat below $m_s^{\rm phys}$.
For $m_x$ we take $n=1,2,3,4$, while for $m_y$ we use $n=8,9,10$.

Our chiral extrapolations are done not with the BSM B-parameters themselves,
but instead with
the ``gold-plated'' ratios introduced in Ref.~\cite{Bailey:2012wb}:
\begin{equation}
G_{23} \equiv \frac{B_2}{B_3}\,, \quad
G_{45} \equiv \frac{B_4}{B_5}\,, \quad
G_{24} \equiv B_2 \cdot B_4\,, \quad
G_{21} \equiv \frac{B_2}{B_K}\,.
\end{equation}
These have the advantage of canceling chiral logarithms at NLO,
so that the chiral extrapolations at this
order involve only analytic terms [see Eq.~(\ref{eq:Xfit}) below].

\begin{figure}[h]
\centering
\subfigure[$G_{23}$]{
\includegraphics[width=0.48\textwidth]{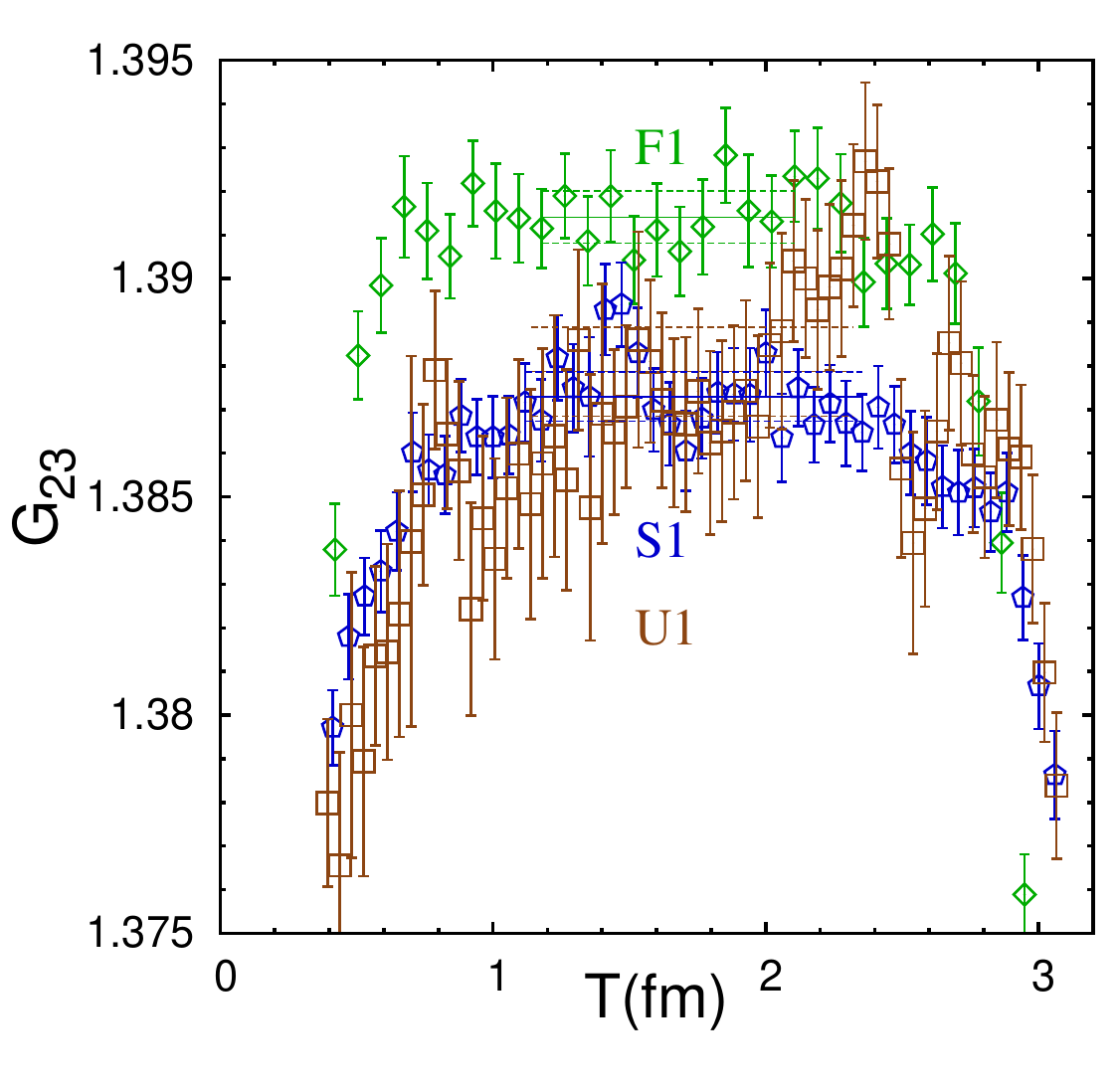}}
\subfigure[$G_{45}$]{
\includegraphics[width=0.48\textwidth]{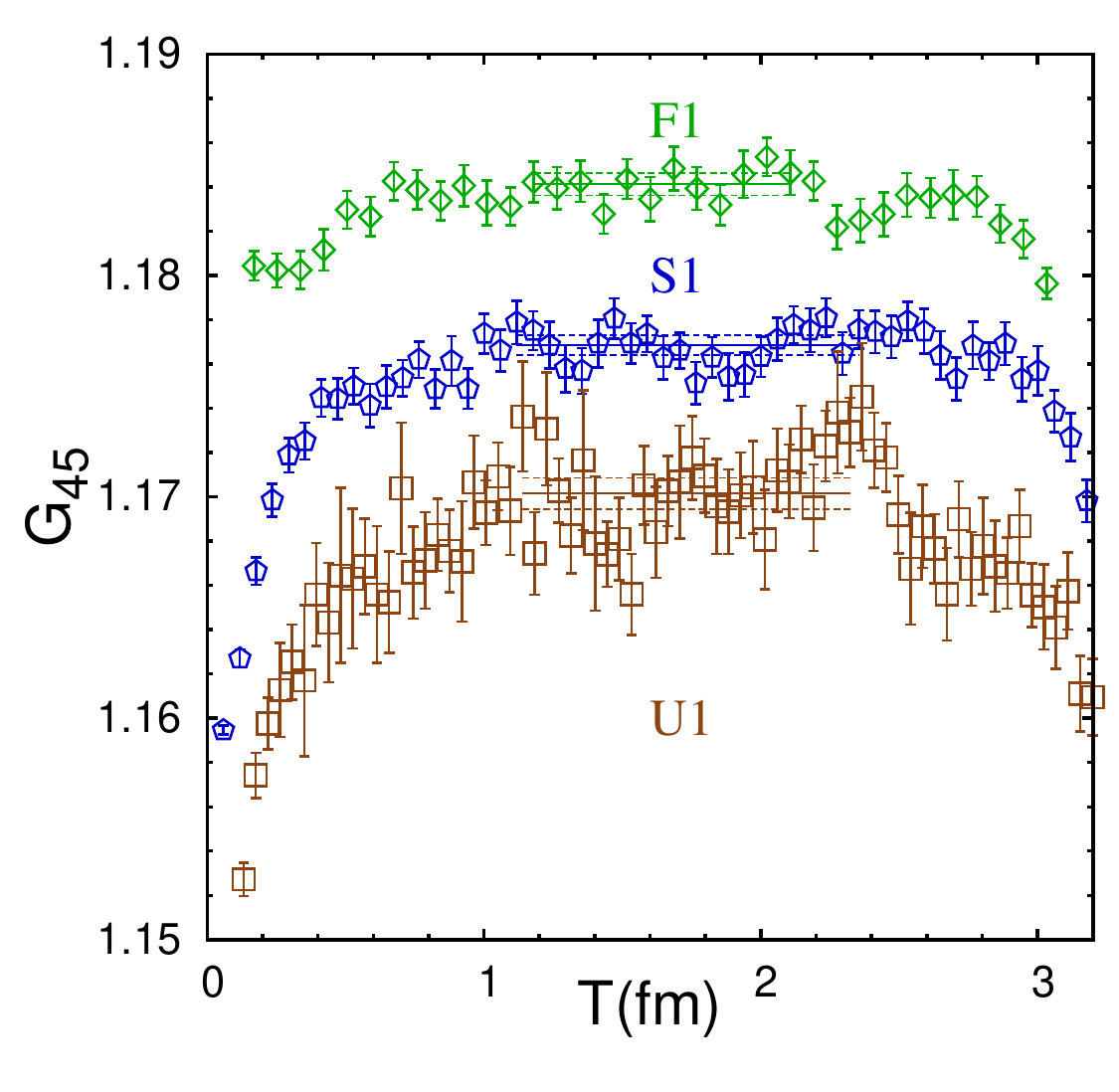}}
\caption{$G_{23}$ and $G_{45}$ at renormalization scale $\mu=1/a$ as a 
function of $T$, the Euclidean time between the kaon source 
and operator insertion. 
(Green) diamond, (blue) pentagon, and (brown) square 
points are from the F1, S1, U1 ensembles, respectively,
which all have similar sea-quark masses ($am_\ell/am_s = 1/5$).
The valence quarks satisfy $(m_x,m_y)=(m_s/10, m_s)$.} 
\label{fig:raw}
\end{figure}

We use U(1) noise sources to create kaon states with a fixed Euclidean
time separation ($\Delta T\approx 3.3$ fm), 
and measure the G-parameters between them.
In Fig.~\ref{fig:raw}, we show representative results for
$G_{23}$ and $G_{45}$.
Here the operators have been matched to the continuum at
the scale $\mu=1/a$.
We observe good plateaus in both quantities.
We also find that the dependence on $a$ is weaker than
for the B-parameters themselves.

On each ensemble, we first do the valence chiral extrapolation. 
This ``X-fit'' is done
with respect to $X \equiv X_p / \Lambda^2_\chi$,
with $\Lambda_\chi =1 \GeV$ and
$X_P =m^2_\pi(x\bar{x})$ the valence $x\bar{x}$ meson mass-squared.
For the G-parameters, the fitting function is
\begin{equation}
G_i(\text{X-fit}) =c_1 + c_2 X + c_3 X^2 + c_4 X^2 \text{ln}^2X 
		  + c_5X^2 \text{ln} X + c_6 X^3\,,
\label{eq:Xfit}
\end{equation}
which includes generic NNLO chiral logarithmic terms.
We do correlated fits with Bayesian priors $c_i = 0 \pm 1$ for $c_{4-6}$.
In the case of $B_K$, the fitting functional includes NLO chiral
logs, and we follow the same fitting procedure as in 
Refs.~\cite{Bae:2010ki,Bae:2011ff,Bae:2014sja}.

\begin{figure}[h]
\subfigure[X-fit]{
  \includegraphics[width=0.48\textwidth]
                  {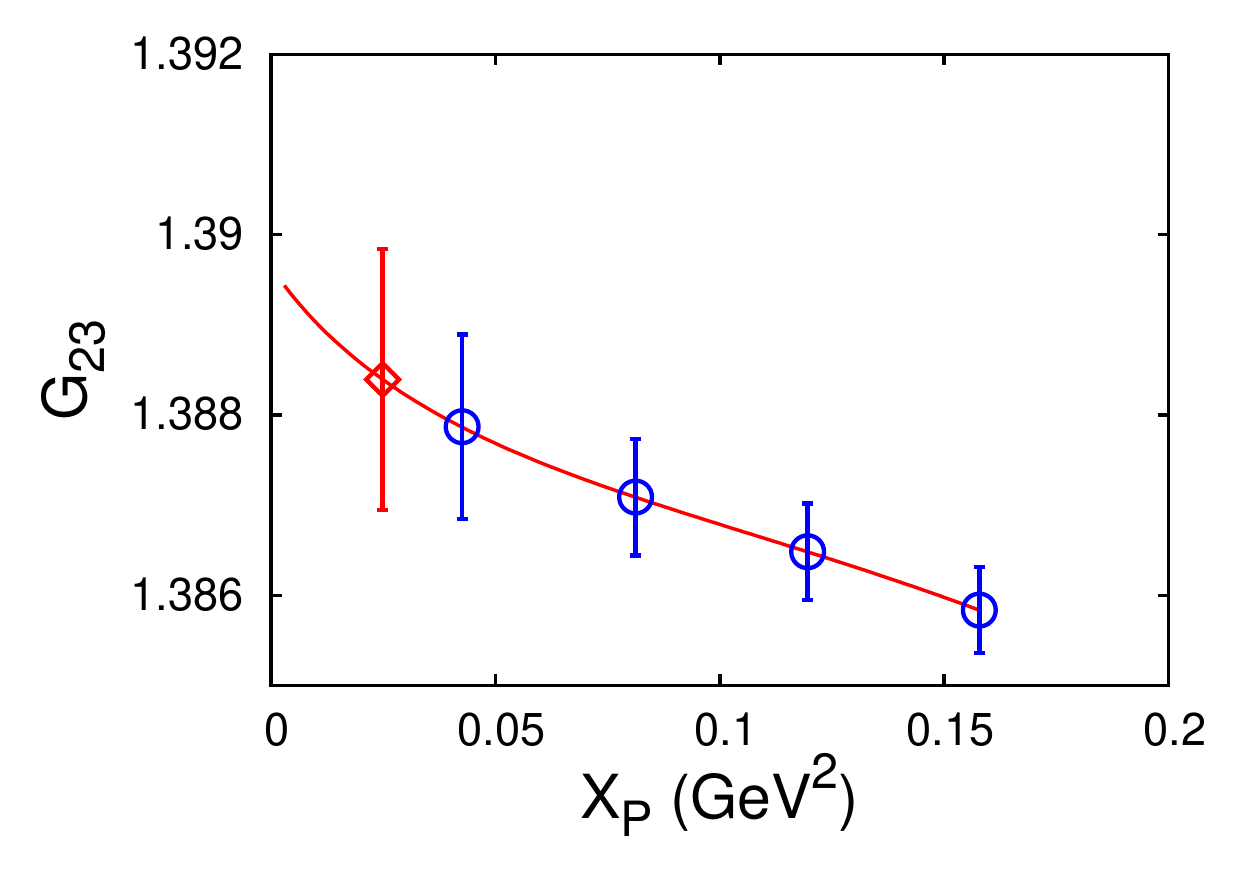}
}
\subfigure[Y-fit]{
  \includegraphics[width=0.48\textwidth]
                  {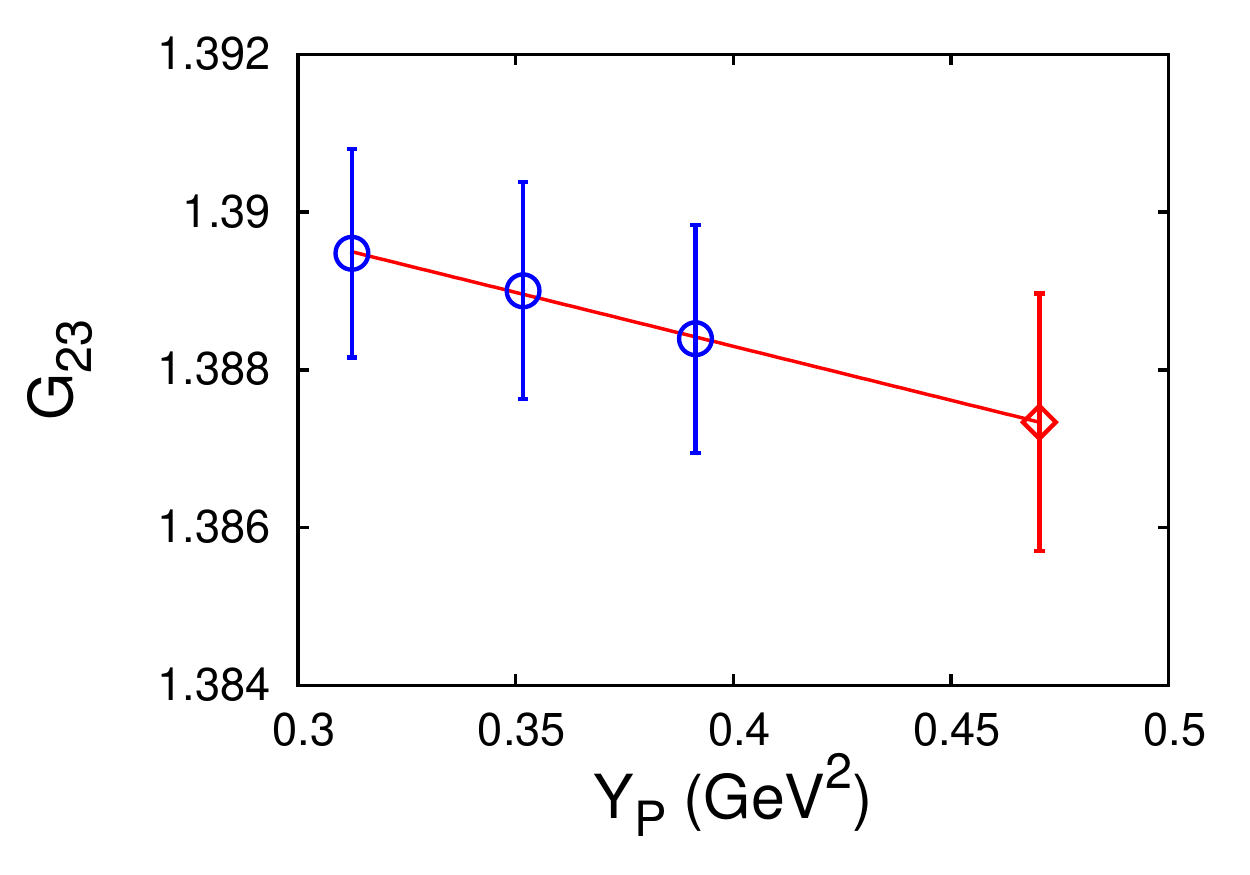}
}
\caption{Extrapolations of $G_{23}$ on the U1 ensemble, 
for operators renormalized at the scale $\mu=1/a$. 
The red points are extrapolated values.}
\label{fig:fitting}
\end{figure}

We next do the ``Y-fit'': an extrapolation in
$Y \equiv Y_P / \Lambda^2_\chi$, with $Y_P = m^2_\pi(y\bar{y})$,
to the ``physical'' $\eta_s(s\bar{s})$ mass.
The central value is obtained by a linear fit.
%
%
Example X- and Y-fits are shown in Fig.\ref{fig:fitting}.

The final step is to use all our ensembles to extrapolate
to the physical values of the sea-quark masses and to the continuum limit.
To do so, we must first evolve the matrix elements (or ratios)
from $\mu=1/a$ to a common energy scale,
$\mu=2\GeV$ or $3\GeV$, 
using the two-loop anomalous dimension matrix from Ref.~\cite{Buras:2000if}.
We have discovered that, in Refs.~\cite{Bae:2013mja,Bae:2013tca}, 
we used the wrong value of the two-loop contribution to the
anomalous dimension for the pseudoscalar operator.
This enters through the denominators of the B-parameters
[see Eq.~(\ref{eq:BSMBdef})]. This error turns out to have a $\sim 5\%$
effect on the final results for BSM B-parameters.
We have corrected this problem here.
\begin{table}[h!]
\begin{center}
\begin{tabular}{| c | c | c| }
\hline\hline
fit type & fitting functional form & Bayesian Constraints \\
\hline
$F^1_B$ & $d_1 + d_2 \frac{L_P}{\Lambda^2_\chi}
        + d_3 \frac{S_P}{\Lambda^2_\chi} 
        + d_4 (a \Lambda_Q)^2$
        & $d_2 \cdots d_4 = 0 \pm 2$ \\
\hline
%
%
$F^4_B$ & $F^1_B + d_5 (a \Lambda_Q)^2 \frac{L_P}{\Lambda^2_\chi}
        + d_6(a \Lambda_Q)^2 \frac{S_P}{\Lambda^2_\chi}$
        & \\
        & $+ d_7 (a \Lambda_Q)^2 \alpha_s 
        + d_8 \alpha_s^2 +d_9 (a \Lambda_Q)^4$
        & $d_2 \cdots d_9 = 0 \pm 2$ \\
\hline
$F^6_B$ & $F^4_B + d_{10} \alpha_s^3 
        + d_{11} (a \Lambda_Q)^2 \alpha_s^2 
        + d_{12} (a \Lambda_Q)^4 \alpha_s 
        + d_{13} (a \Lambda_Q)^6 $
        & \\
        & $+ d_{14}(a \Lambda_Q)^4 \frac{L_P}{\Lambda^2_\chi} 
        + d_{15}(a \Lambda_Q)^2\alpha_s\frac{L_P}{\Lambda^2_\chi} 
        + d_{16}\alpha_s^2 \frac{L_P}{\Lambda^2_\chi}$ & \\
        & $+ d_{17}(a \Lambda_Q)^4 \frac{S_P}{\Lambda^2_\chi} 
        + d_{18}(a \Lambda_Q)^2\alpha_s\frac{S_P}{\Lambda^2_\chi} 
        + d_{19}\alpha_s^2 \frac{S_P}{\Lambda^2_\chi}$
        & $d_2 \cdots d_{19} = 0 \pm 2$ \\
\hline\hline
\end{tabular}
\caption{Fitting functional forms for continuum-chiral extrapolation.}
\label{tab:fit-func}
\end{center}
\end{table}
For the continuum-chiral extrapolation we use the fitting forms
listed in Table \ref{tab:fit-func}.
These are based on the power-counting rules for SU(3) SChPT,
i.e. $L_P/\Lambda^2_{\chi} \approx
S_P/\Lambda^2_{\chi} \approx (a\Lambda_Q)^2 \approx \alpha_s$,
with $L_P$, $S_P$ the squared masses of the sea-quark pion and
$\bar ss$ meson, respectively.
Fit-form $F^1_B$ contains terms up to NLO, with the $\alpha_s$ term
absent since we use 1-loop perturbative matching.  $F^4_B$ contains
all NNLO terms except those quadratic in $L_P$ and $S_P$.
$F^6_B$ contains all NNNLO terms which are up to linear in quark masses.

\begin{figure}[htb]
\subfigure[ $G_{24}$ fit by $F^1_B$, $\chi^2/\text{dof} = 1.08$]
{\includegraphics[width=.48\textwidth]{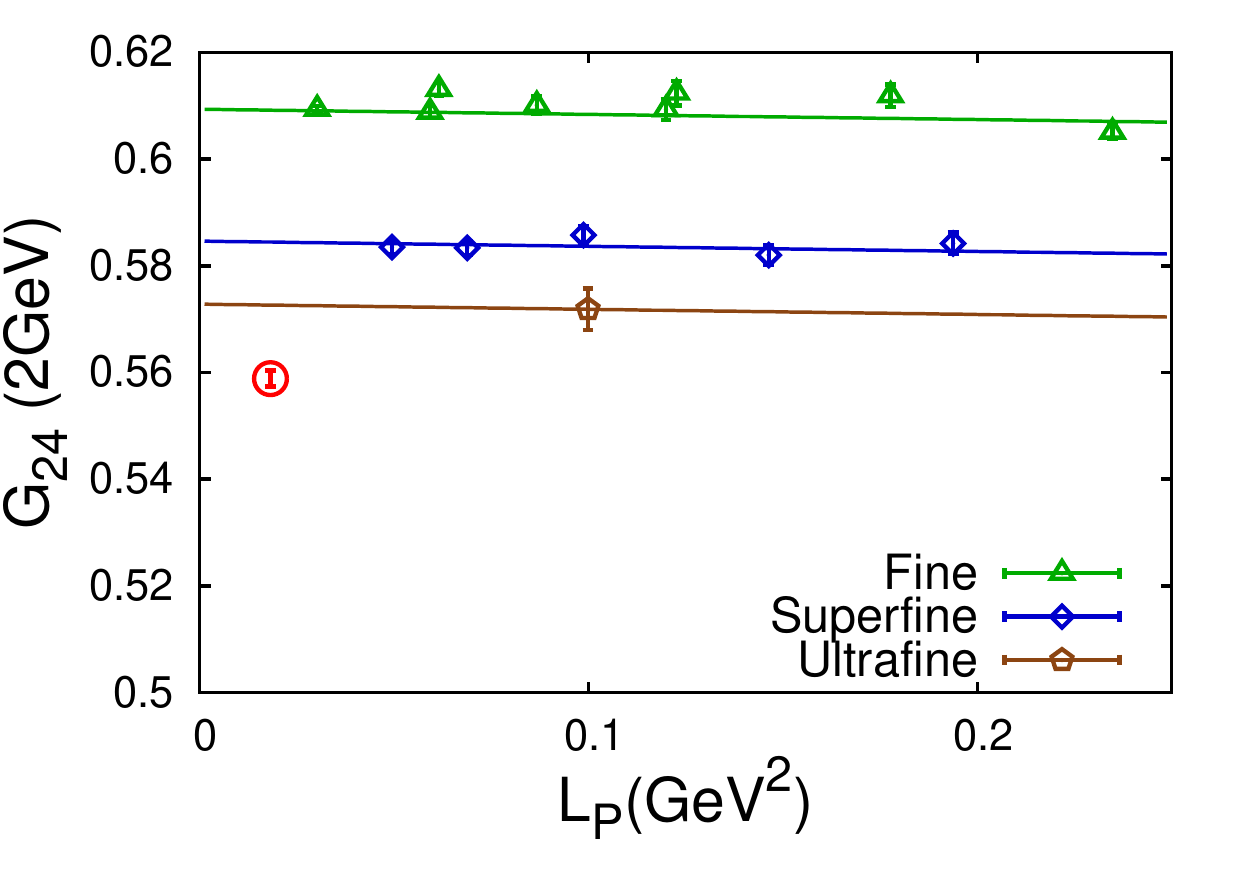}}
\subfigure[ $G_{24}$ fit by $F^4_B$, $\chi^2/\text{dof} = 0.91$]
{\includegraphics[width=.48\textwidth]{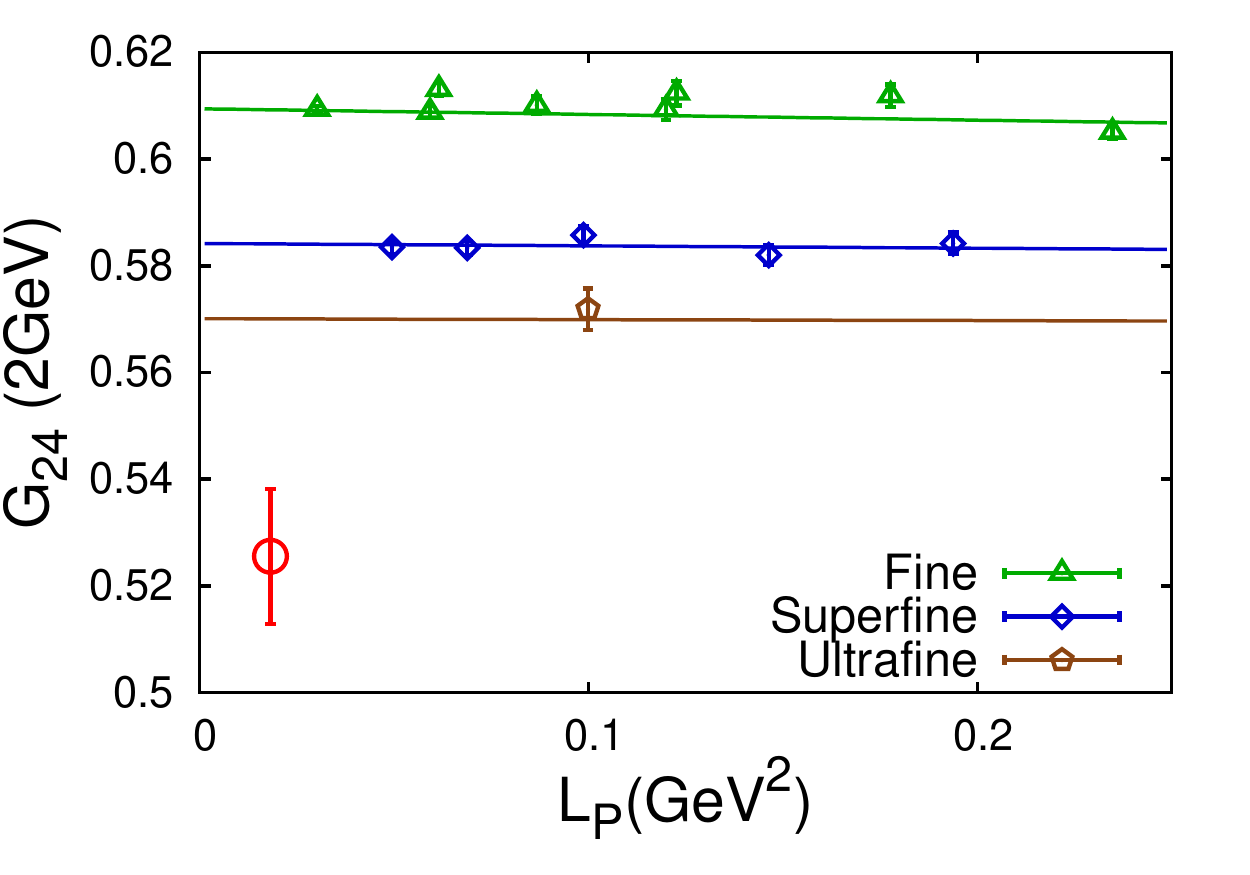}}
\caption{Continuum-chiral extrapolation of $G_{24}$.  Red circles are
extrapolated results. Since values of $1/a$ and $S_P$ vary among ensembles
with the same nominal lattice spacing, some scatter about the horizontal lines
is expected, and this is accounted for in the fits [though not in the plot].}
\label{fig:cont-chiral:G_24}
\end{figure}

\begin{figure}[htb]
\label{fig:cont-chiral}
\subfigure[ $G_{45}$ fit by $F^1_B$ , $\chi^2/\text{dof} = 4.06$]
{\includegraphics[width=.48\textwidth]{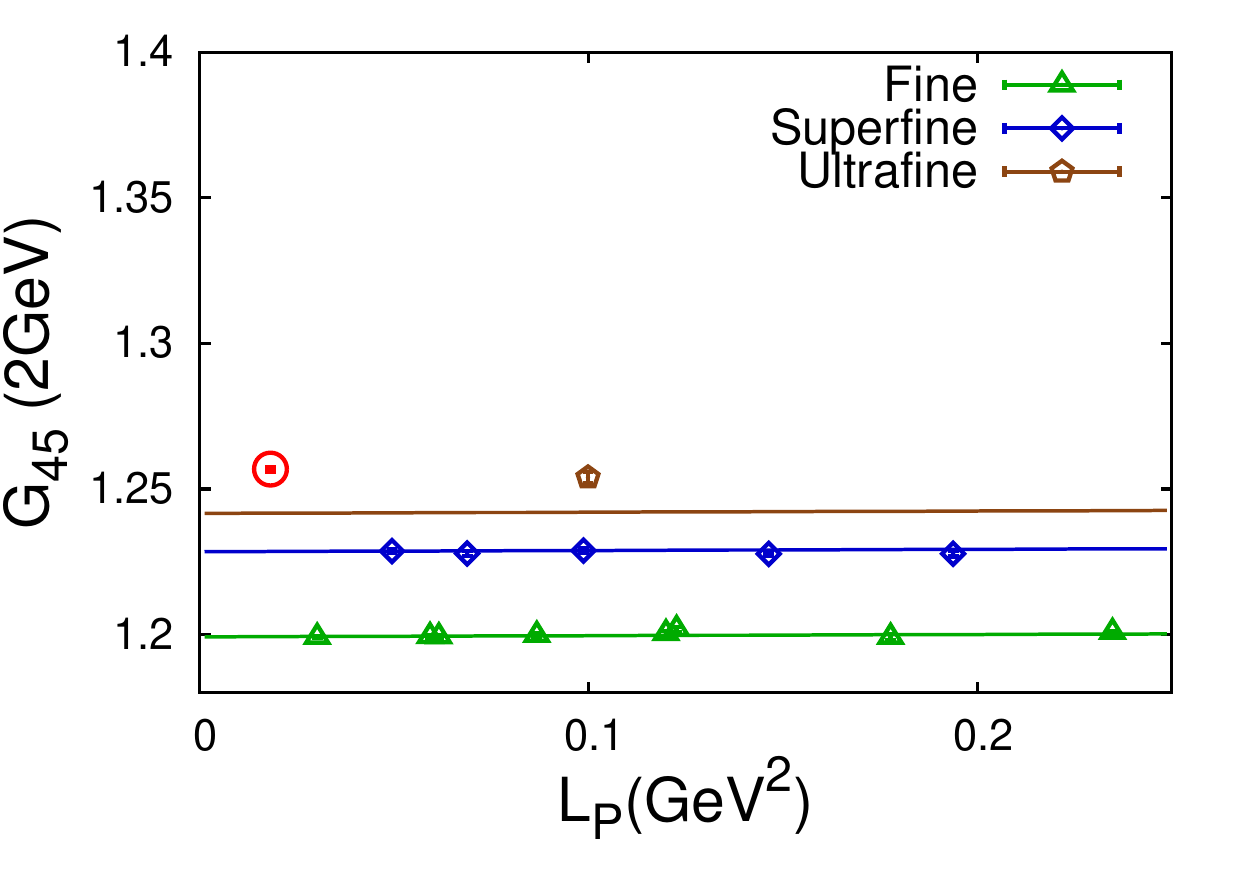}}
\subfigure[ $G_{45}$ fit by $F^4_B$ , $\chi^2/\text{dof} = 1.38$]
{\includegraphics[width=.48\textwidth]{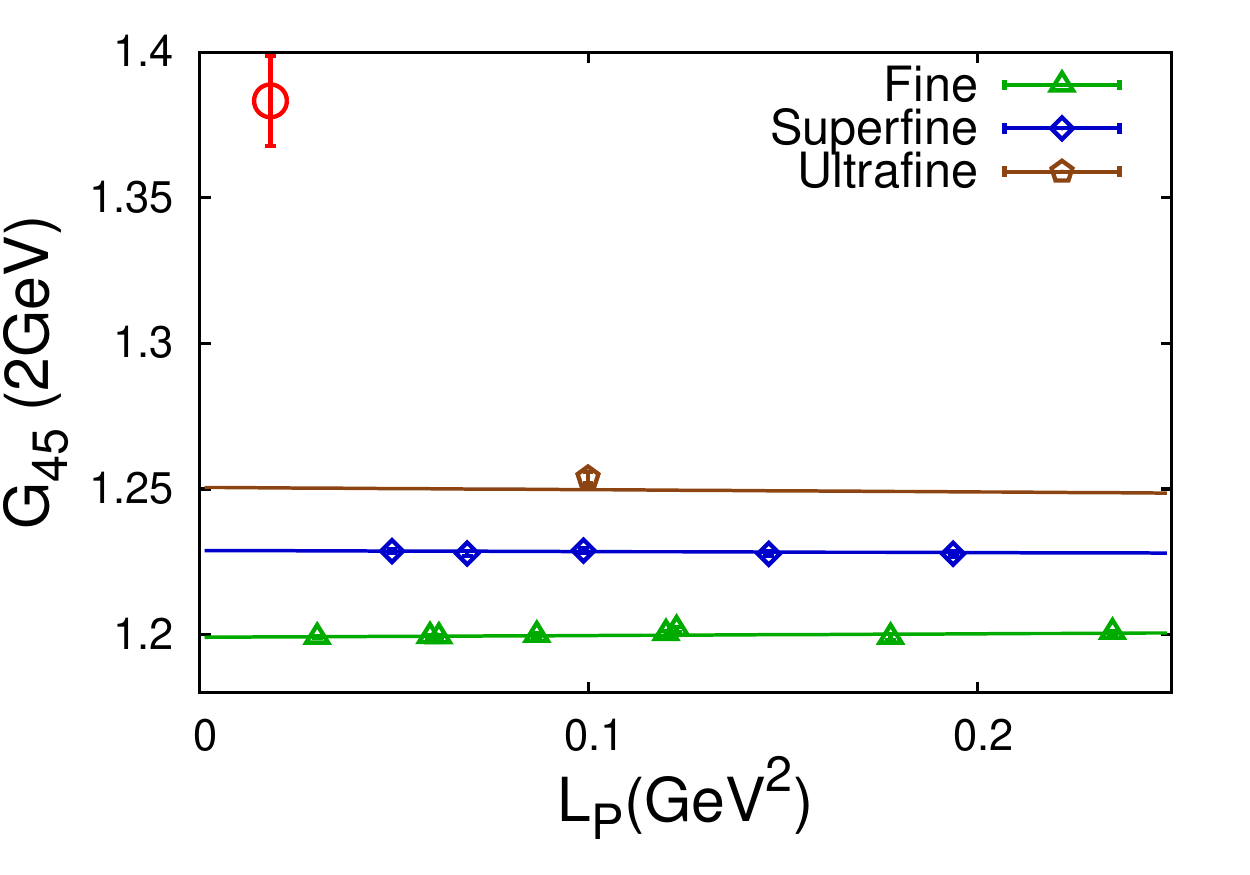}} \\
\caption{Continuum-chiral extrapolation of $G_{45}$. 
Notation as in Fig.~\protect\ref{fig:cont-chiral:G_24}.}
\label{fig:cont-chiral:G_45}
\end{figure}

In our previous work we used the simplest fitting form,
$F^1_B$, to determine the central values of the
extrapolated result~\cite{Bae:2013mja,Bae:2013tca}.
This continues to work well for
$B_K$, $G_{21}$ and $G_{24}$, giving fits with
$\chi^2/\text{dof} = 1.08 \sim 1.25$, as illustrated in 
Fig.~\ref{fig:cont-chiral:G_24}(a).
However, for $G_{23}$ and $G_{45}$, we find the addition of new ensembles 
leads to $F^1_B$ giving poor fits, and so we use
fit $F^4_B$ instead for our central values.
This gives reasonable fits, with $\chi^2/\text{dof} = 1.32 \sim 1.38$,
as illustrated in Fig.~\ref{fig:cont-chiral:G_45}.

We take the difference between $F_1^B$ and $F_4^B$ as the systematic
uncertainty coming from the continuum-chiral extrapolation.
It is a $\sim 1-8\%$ error. 
We compare it with our earlier estimate of the systematic error from
one-loop perturbative matching ($\alpha_s^2(U1)\approx 4.4$\%), and
quote the larger as our estimate of the systematic error.
%
%
%
%
%
%


\section{Discussion of results}
The changes to our results since Lattice 2013 are 
summarized in Table~\ref{tab:change}.
The first column shows the results presented last year in 
Refs.~\cite{Bae:2013mja,Bae:2013tca}.
The impact of adding new ensembles is shown in the second column,
labeled ``2014(ens)'', and is minor.
The effect of correcting the two-loop pseudoscalar anomalous dimension
is shown in the third column, labeled ``2014(A.D.)''.
This leads to $\sim 5\%$ reductions in the BSM B-parameters.
Up to this stage the results are from $F_1^B$ fits.
The final column shows the impact of switching to $F^4_B$ fits
for $G_{23}$ and $G_{45}$ (necessitated by the poor quality
of the $F_1^B$ fits). This changes $B_3$ 
and $B_5$ by $\sim 3\%$ and $9\%$, respectively.

\begin{table}[tbp]
\begin{center}
\begin{tabular}{| c | c  |c |c | c| }
\hline\hline
B($\mu=3$GeV)              & 2013  & 2014(ens) & 2014(A.D.) & 2014(final) \\
\hline
$B_K$                   &0.519(7)(23)&  0.518(3) &0.518(3)&0.518(3)(23)\\
$B_2$                   &0.549(3)(28)&  0.547(1) &0.525(1)&0.525(1)(23)\\
$B_3$    		&0.390(2)(17)&  0.390(1) &0.375(1)&0.358(4)(18)\\
$B_3^{\text{SUSY}}$     &0.790(30)   &  0.783(2) &0.750(2)&0.774(6)(64)\\
$B_4$                   &1.033(6)(46)&  1.024(1) &0.981(3)&0.981(3)(61)\\
$B_5$                   &0.855(6)(43)&  0.853(3) &0.817(2)&0.748(9)(79)\\
\hline\hline
\end{tabular}
\caption{
Changes to results during the last year. See text for
discussion. Only statistical errors are shown in the second and third columns.}
\label{tab:change}
\end{center}
\end{table}
In Table~\ref{tab:compare} we show (in the first two columns) 
our preliminary results
for two choices of renormalization scale $\mu$.
The dominant error in the BSM B-parameters
comes from the chiral-continuum extrapolation.
For further details on the error budget, see 
Refs.~\cite{Bae:2013tca,Bae:2014sja}.
\begin{table}[htbp]
\begin{center}
\begin{tabular}{| l | c ||c | c|  c| }
\hline\hline
                        & \multicolumn{2}{c|}{SWME}  & RBC\&UKQCD & ETM \\
\cline{2-5}
                        & $\mu=2$GeV    & $\mu=3$GeV & $\mu=3$ GeV & $\mu=3$ GeV\\
\hline
$B_K$                   &0.537(4)(24) &0.518(3)(23) &0.53(2)&0.51(2)\\
$B_2$                   &0.568(1)(25) &0.525(1)(23) &0.43(5)&0.47(2)\\
$B_3$    		&0.380(4)(19) &0.358(4)(18) &N.A.   &N.A.   \\
$B_3^{\text{SUSY}}$     &0.849(6)(69) &0.774(6)(64) &0.75(9)&0.78(4)\\
$B_4$                   &0.984(3)(63) &0.981(3)(61) &0.69(7)&0.75(3)\\
$B_5$                   &0.712(9)(80) &0.748(9)(79) &0.47(6)&0.60(3)\\
\hline\hline
\end{tabular}
\caption{Comparison of results with those of 
RBC/UKQCD~\cite{Boyle:2012qb} and ETM~\cite{Bertone:2012cu} collaborations.}
\label{tab:compare}
\end{center}
\end{table}
%
We also compare our results with those published by other collaborations.
Results for $B_K$ and $B_3^{\rm SUSY}$ are consistent, those
for $B_2$ differ at $\sim 2\sigma$, and those for $B_4$ and $B_5$
differ at $\sim 3\sigma$.
These discrepancies, have, however, been reduced during the last year due
to the changes summarized in Table~\ref{tab:change}.

The source of these discrepancies is not yet clear.
We see two (related) places where we can improve
our understanding of the systematic errors. 
First, we are using one-loop matching, whereas the other collaborations
use non-perturbative renormalization (NPR). Our error estimate assumes a
two-loop term of relative size $\alpha_s^2$, and this could be an
underestimate.
We are working towards using NPR to normalize our operators.
Second, our continuum extrapolation is not fully satisfactory,
as exemplified by the results in Fig.~\ref{fig:cont-chiral:G_45}. 
The ultrafine ensemble lies further from the superfine and fine ensembles
than is consistent with the simple $F^1_B$ fit-form, and
this drives the fit in the right-hand panel to pick out relatively large
coefficients for $\alpha_s^2$ terms. While this may be correct,
it leads to a non-intuitive extrapolation.
We hope to report further on these issues in the near future.

\section{Acknowledgments}
We are grateful to Claude Bernard and the MILC collaboration
for private communications.
C.~Jung is supported by the US DOE under contract DE-AC02-98CH10886.
The research of W.~Lee is supported by the Creative Research
Initiatives Program (No.~2014001852) of the NRF grant funded by the
Korean government (MEST).
W.~Lee would like to acknowledge the support from KISTI supercomputing
center through the strategic support program for the supercomputing
application research (KSC-2013-G2-005).
The work of S.~Sharpe is supported in part by the US DOE grants
no.~DE-FG02-96ER40956 and DE-SC0011637.
%
%
%
\bibliographystyle{JHEP}
\bibliography{ref}

\providecommand{\href}[2]{#2}\begingroup\raggedright\begin{thebibliography}{10}

\bibitem{Bae:2013mja}
T.~Bae et~al. {\em PoS} {\bf LATTICE2013} (2013) 473,
  [\href{http://xxx.lanl.gov/abs/1310.7372}{{\tt arXiv:1310.7372}}].

\bibitem{Bae:2013tca}
T.~Bae et~al. {\em Phys.Rev.} {\bf D88} (2013) 071503,
  [\href{http://xxx.lanl.gov/abs/1309.2040}{{\tt arXiv:1309.2040}}].

\bibitem{Boyle:2012qb}
P.~Boyle, N.~Garron, and R.~Hudspith {\em Phys.Rev.} {\bf D86} (2012) 054028,
  [\href{http://xxx.lanl.gov/abs/1206.5737}{{\tt arXiv:1206.5737}}].

\bibitem{Bertone:2012cu}
V.~Bertone et~al. {\em JHEP} {\bf 1303} (2013) 089,
  [\href{http://xxx.lanl.gov/abs/1207.1287}{{\tt arXiv:1207.1287}}].

\bibitem{Buras:2000if}
A.~J. Buras, M.~Misiak, and J.~Urban {\em Nucl.Phys.} {\bf B586} (2000)
  397--426, [\href{http://xxx.lanl.gov/abs/hep-ph/0005183}{{\tt
  hep-ph/0005183}}].

\bibitem{Gabbiani:1996hi}
F.~Gabbiani, E.~Gabrielli, A.~Masiero, and L.~Silvestrini {\em Nucl.Phys.} {\bf
  B477} (1996) 321--352, [\href{http://xxx.lanl.gov/abs/hep-ph/9604387}{{\tt
  hep-ph/9604387}}].

\bibitem{Bailey:2012wb}
J.~Bailey, H.-J. Kim, W.~Lee, and S.~Sharpe {\em Phys.Rev.} {\bf D85} (2012)
  074507, [\href{http://xxx.lanl.gov/abs/1202.1570}{{\tt arXiv:1202.1570}}].

\bibitem{Bae:2010ki}
T.~Bae et~al. {\em Phys.Rev.} {\bf D82} (2010) 114509,
  [\href{http://xxx.lanl.gov/abs/1008.5179}{{\tt arXiv:1008.5179}}].

\bibitem{Bae:2011ff}
T.~Bae et~al. {\em Phys.Rev.Lett.} {\bf 109} (2012) 041601,
  [\href{http://xxx.lanl.gov/abs/1111.5698}{{\tt arXiv:1111.5698}}].

\bibitem{Bae:2014sja}
T.~Bae et~al. {\em Phys.Rev.} {\bf D89} (2014) 074504,
  [\href{http://xxx.lanl.gov/abs/1402.0048}{{\tt arXiv:1402.0048}}].

\end{thebibliography}\endgroup

\end{document}